# A new low-cost technique improves weather forecasts across the world

*Tim D. Hewson[1] and Fatima M. Pillosu[2],
[1]ECMWF, Reading, UK
[2]ECMWF and The University of Reading, UK

## *Abstract*

Computer-generated forecasts divide the earth's surface into gridboxes, each now ~25% of the size of London, and predict one value per gridbox. If weather varies markedly within a gridbox forecasts for specific sites inevitably fail. A completely new statistical post-processing method, using ensemble forecasts as input, anticipates two gridbox-weather-dependant factors: degree of variation in each gridbox, and bias on the gridbox scale. Globally, skill improves substantially; for extreme rainfall, for example, useful forecasts extend 5 days ahead. Without post-processing this limit is <1 day. Relative to historical forecasting advances this constitutes ground-breaking progress. The key drivers, incorporated during calibration, are meteorological understanding and abandoning classical notions that only local data be used. Instead we simply recognise that "showers are showers, wherever they occur worldwide" which delivers a huge increase in calibration dataset size. Numerous multi-faceted applications include improved flash flood warnings, physics-related insights into model weaknesses and global pointwise re-analyses.

Weather forecasts nowadays rely heavily on computer-based models, i.e. numerical weather prediction (NWP)[1], and commonly an ensemble of predictions is used, to represent uncertainties[2]. Due to computational power limitations, a gridbox in the best operational global ensembles currently spans about 20 km by 20 km in the horizontal (hereafter: "GM scales" = Global Model scales). So NWP forecasts do not output rainfall (for example) at specific sites, that most customers require, but instead "average rainfall" for much larger gridboxes. This disconnect is an important forecasting problem, which this study addresses. To elaborate, we introduce here the notion of "sub-grid variability", to mean the variation seen amongst all point values observed within the same model gridbox. If sub-grid variability is low then raw NWP forecasts can provide accurate forecasts for points. But if sub-grid variability is high such forecasts inevitably fail.

The most common strategies to address sub-grid variability problems are using a much higher resolution model (e.g. ~2 km*2 km) to minimise them[3], or using calibrated post-processing (PP) techniques to statistically convert from gridbox to point forecasts[4–6]. For predicting rainfall, the parameter central to this article, high resolution models, whilst showing much more realistic-looking spatial patterns, exhibit rather limited improvements in forecast skill[7,8]. And because of computational constraints one such model might only cover 0.2 % of the world. For global coverage PP techniques are a better prospect, and they have historically performed well in improving forecasts of dry weather[6,9], but as previous authors themselves acknowledge, in critiques of their own and others' work, those techniques can present a number of challenges and issues (Table 1). Table 1 also highlights how our brand-new approach, described in this study, addresses these points. Ours is a non-local gridbox-analogue approach, formulated via the principles of conditional verification[10], with some structural similarities to quantile regression forests[11,12]. We call the method "ecPoint" - "ec" for the European Centre (for Medium-Range Weather Forecasts), "Point" for point forecasts.

**Table 1**: **What (previously reported) challenges/issues do classical post-processing methods often face?**

| | Challenge/Issue* | Related characteristics of ecPoint |
|---|---|---|
| 1 | Calibration requires ≳20 years of observations[5,6,13–17] | Vast training datasets come from just 1 year of data |
| 2 | † Calibration requires ≳20 years of ensemble re-forecasts[6,15,17,18] | 1 year of re-forecasts, from just one (Control) run, for up to 48h lead times, is sufficient |
| 3 | Lack of climatological stationarity potentially a problem[16,17,19,20] | 1 year of training is short enough to avoid this |
| 4 | Forecasts not possible where no training data is available[17] | Forecasts can be created for all locations |
| 5 | Distribution fitting often used may not mirror real data[5,6,12,21] | Nonparametric methods are used |
| 6 | Distribution fitting struggles to represent tails well[5,6,16,21,22]([12]) | There are no constraints on tail structure |
| 7 | Difficult to improve forecasts of extremes[5,6,9,12,18,21–25]([26,27]) | Forecasts of 'extremes' are substantially better |
| 8 | Occasional errors in training data may contaminate forecasts[27] | Impacts of such errors are nil or negligible |
| 9 | No pointers regarding "reasons" for model errors / biases([26]) | "Mapping functions" used denote model characteristics |
| 10 | Post-processed spatial output may require smoothing[9,17] | Aesthetically smooth fields arise naturally |
| 11 | Large scale/convective precipitation are not disaggregated([28]) | An intrinsic and valuable facet of the method |
| 12 | ‡ Coverage is not global | Coverage is global |

*Which are relevant depends on the method. Citations outside(inside) parentheses denote where these aspects have been referenced without(with) a mechanism for addressing.
† In the U.S. calibration is the driving force for running re-forecasts (Tom Hamill, personal communication). ECMWF's re-forecast strategy is different.
‡ No cited work has global coverage.

Applications of ecPoint include the many spheres that would benefit from improved probabilistic point forecasts. For rainfall, flash flood prediction is one application, given that we achieve much improved forecasts of localised extremes, as will be shown.

## The concept

Sub-grid variability in rainfall is itself very variable (Fig. 1) and relates closely to the weather situation. There are clear-cut physical reasons for this. Dynamics-driven (large-scale) rainfall, often related to atmospheric fronts, arises from steady ascent of moist air across regions typically larger than GM scales (Fig. 1a). As rainfall rates mirror ascent rates, rainfall rate sub-grid variability tends to be small. Conversely instability-driven rainfall (i.e. showers/convection) arises from localised pockets of rapid ascent, which are typically hundreds of metres to kilometres across. So during convection rainfall rate sub-grid variability, on GM scales, can be very large indeed[29] (Fig. 1b,c).

Rainfall totals arise from integrating rainfall rates over time. Sub-grid variability in totals mostly reduces in proportion to period length. Let us consider, as in many rainfall PP studies, daily to sub-daily time periods (e.g. 6, 12, 24 h), and focus on convection. The intensity, dimensions, density, genesis rate, longevity and speed of movement of convective cells all impact upon the sub-grid variability in totals. For example, cells moving with speed V, that retain intensity and dimensions for a period t, will deliver stripes in a totals field of length V*t. Typical values of V and t might be 15 m/s and 1 h, giving a stripe 54 km long, which is much greater than GM scales. In such situations we thus get sub-grid variability primarily in one dimension (Fig. 1b). In the limiting case of slow-moving cells, where V -> 0, we retain (large) sub-grid variability in two dimensions (Fig. 1c), and sometimes many locations within "wet gridboxes" stay dry.

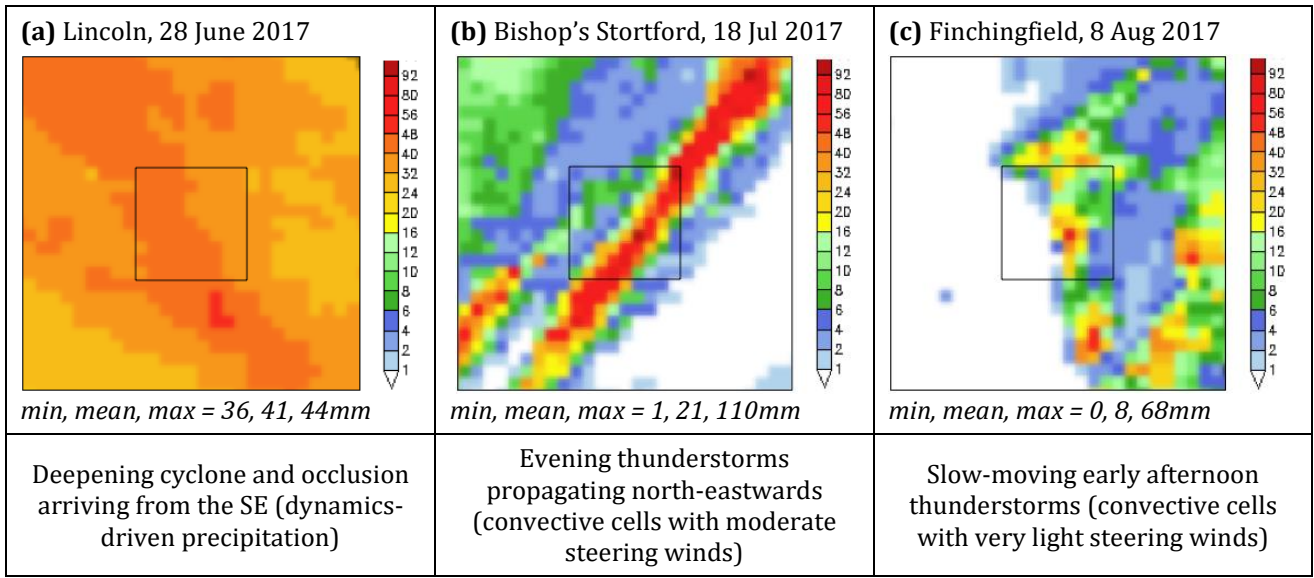

**Fig. 1**: **24 h radar-derived rainfall totals (mm) in the UK showing different types of sub-grid variability. a-c)** 3 cases: coloured rainfall cells measure 2x2 km, full frames 54x54 km. Central black squares denote ECMWF ensemble gridbox size (18x18 km), for which minimum, mean and maximum rainfall is shown beneath. Named locations lie approximately mid-panel; all are in Eastern England, where relatively flat topography makes radar-derived totals more reliable. Row 2 explains the synoptic situation in each case. Images are from *netweather.tv*.

The embodiment of ecPoint is that features of the NWP gridbox forecast output (and other global datasets) can tell us what degree of sub-grid variability to expect. For example, NWP output commonly subdivides rainfall into dynamics-driven and convective, and then for convective cases shower movement speed can be approximated by (e.g.) the 700 hPa wind speed ($V_{700}$). So by using the convective rainfall fraction (CF), and $V_{700}$ (two "governing variables") we can distinguish each of the 3 types on Fig. 1, to anticipate a priori the expected sub-grid variability, and accordingly convert each forecast for each gridbox into a probabilistic point rainfall prediction. To our knowledge this general approach, based on first principles of precipitation generation, has not been used before except in a limited way for nowcasting[28] (Table 1, row 11).

The above logic could be successfully applied to a deterministic forecast, but in NWP ensembles furnish the most useful predictions[2,30]. So instead we apply to ensemble predictions, creating an ensemble of probabilistic realisations (or "ensemble of ensembles") that we merge to give the final probabilistic point forecast.

## Calibration

All PP methods need to be calibrated. For ecPoint we appeal to the concept of conditional verification to create a separate 'mapping function' (***M***) for each of the m possible combinations of governing variable ranges. Each such function aims to represent possible outcomes of point rainfall within a gridbox and so each should represent a different type of sub-grid variability. In our very simple Fig. 1 example there would be 3 categories (A, B, C), which we call the 'gridbox-weather-types' (or 'types'), which would have governing variable characteristics like these:

- A : CF < 0.5
- B : CF > 0.5 and $V_{700}$ > 5m/s
- C : CF > 0.5 and $V_{700}$ < 5m/s

Then mapping functions $M_A$, $M_B$, $M_C$ would represent probability density functions (pdfs) of point rainfall within a gridbox for, respectively, types j = A, B, C. To create, in the general case, m mapping functions we need to allocate, to one of the m types, each and every rain gauge observation ($r_0$) taken during a pre-defined period, ideally 1 year, and over a pre-defined region, ideally the world. At the same time, we must relate these to forecast gridbox rainfall totals (G) that inevitably differ, and to do this we introduce a nondimensional metric, the "forecast error ratio" (FER):

$$\text{FER} = (r_0 - G)/G \qquad \text{(requires G} \geq \text{1mm)} \qquad (1)$$

$$\text{So that} \quad M = M(x = \text{FER}) \quad \text{with} \quad \int_{-1}^{\infty} M \, dx = 1$$

Naturally we assign each $r_0$ value to a gridbox based on location and assign a companion type to that gridbox at that time using values of the selected governing variables. Short range (unperturbed) control run forecasts provide these values, and $G = G_{control}$. Then, after discarding cases with $G_{control}$<1mm for stability/discretization reasons, each remaining ($G_{control}$, $r_0$) pair furnishes one FER value for the said gridbox-weather-type. By accumulating these we ultimately generate one FER pdf for each mapping function, i.e. $M_1$, ... $M_j$, ... $M_m$. These form the calibration procedure output; 5 examples are shown on Figs. 2a-e. Equation 1 indicated that NWP "over-prediction" for a rain gauge site was represented by FER < 0 (see dark green and green bars), and "under-prediction" by FER > 0 (see yellow and red bars).

On Fig. 2 examples sub-grid variability magnitude depends strongly on the gridbox-weather-type. Figures 2a-c, for example, broadly correspond, respectively, to examples in Figs. 1a-c. Variability is lowest in 2a (distribution roughly Gaussian) and highest, with the greatest likelihood of zeros, in 2c (distribution roughly exponential). This correspondence supports the use of worldwide gauge observations for our "non-local" calibration. In standard convective situations (Figs. 2b,c) "good forecasts" (white bar) are evidently rare, whilst substantial under-prediction (red), which could lead to "unexpected" flash floods, is relatively common, especially when steering winds are light (Fig. 2c).

Physical reasoning and case studies suggest that *gridscale bias* in raw NWP forecasts can also depend strongly on gridbox-weather-type. As well as predicting sub-grid variability ecPoint also corrects for this bias. The gridscale bias correction factor ($C_j$) is implicit in each mapping function; it derives from the "expected value" of the associated FER pdf:

$$C_j = 1 + \int_{-1}^{\infty} x M_j \, dx \qquad (2)$$

$$\approx (\text{mean rainfall observed})/(\text{mean rainfall forecast}) \quad \text{for type } j$$

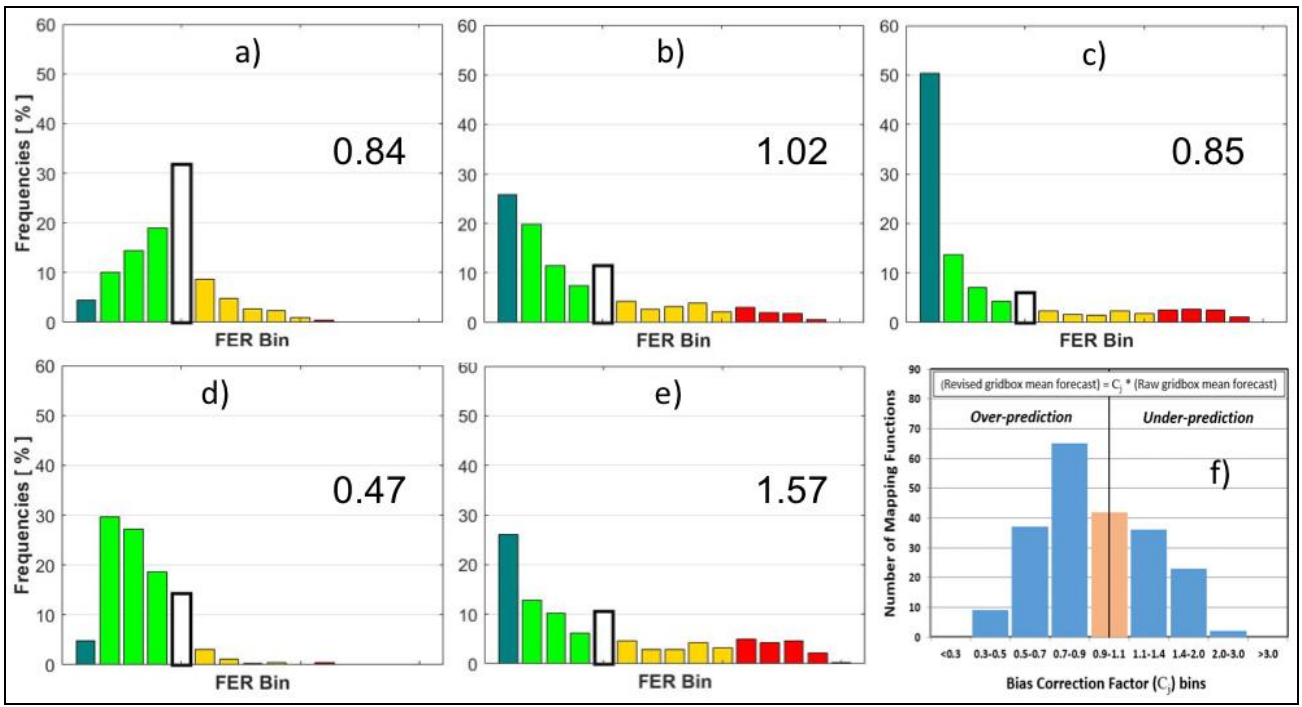

**Fig. 2: Mapping function examples, and the bias correction factor distribution**. **a-e)** Mapping function examples from current operational forecast system shown as histograms; large numbers show implicit bias correction factors C. Dark green, green, white, yellow and red bars denote, respectively, FER ranges for "mostly dry" (<-0.99), "over-prediction" (-0.99 to -0.25), "good forecasts" (-0.25 to 0.25), "under-prediction" (0.25 to 2) and "substantial under-prediction" (2 to ∞). Bar and colour boundaries were subjectively chosen. **a)** is for one weather type with >75 % large-scale rainfall. **b)** for >75% convective rainfall, moderate CAPE (convective available potential energy), strong steering winds, **c)** for >75 % convective, moderate CAPE, light winds, **d)** for >75 % convective, large forecast totals, low CAPE (= weather type "44210" - see Fig. 5b,c), **e)** for 25–50 % convective, modest forecast totals, moderate winds, moderate CAPE. **f)** C value distribution for all 214 functions.

Figure 2f shows the range of $C_j$ values for all mapping functions currently used; for some weather types the ECMWF model seems to markedly over- or under-predict rainfall (e.g. Figs. 2d and e respectively). This is key for ecPoint and is informative also for forecasters and model developers. Although not a definitive inference, Fig. 2f supports the working assumption[31] that *on average* the ECMWF model over-predicts rainfall. The equivalence on line 2 of Eqn. 2 assumes $\partial C_j / \partial G \cong 0$. The larger the range of G values the less likely this is to be valid. We reduce ranges by using G as a governing variable for each type *j*, thereby increasing the efficacy of our model bias interpretation.

A feature of our method is the freedom for the user to select, test and incorporate any variable that can influence rainfall sub-grid variability and/or bias. The following variable "classes" are included: "*raw model*", "*computed*", "*geographical*", "*astronomical*". The first two relate directly to NWP output, the second two to other datasets (see Methods). In Methods we also discuss assumptions implicit in the approach, and the semi-objective strategy for defining governing variable breakpoints.

Unlike classical methods, location is not used in the calibration. This concurs with tests of site separation importance that allocate only 2% weight to this factor[32]. Removing location facilitates the generation of immense training datasets (matching recommendations[17]) that deliver on average ~$10^4$ cases for each mapping function. An example will illustrate the powerful implications of this for ecPoint. Consider a Swedish site experiencing a gridbox-weather-type that is a one-in-five-year event there (e.g. locally extreme Convective Available Potential Energy, i.e. CAPE). Such types do occur much more often globally, enough to deliver

~$10^4$ calibration cases for that type in 1 year of global data. Handling this is computationally straightforward, and moreover the huge case count facilitates our hyper-flexible non-parametric approach. To match this in a simple deterministic MOS approach[4], or a state-of-the-art ensemble approach with 50 supplementary sites[33], would conversely require the impossible - i.e. data for the last 50,000 or 1,000 years respectively.

## Forecast Methodology

ecPoint forecast production relies on converting eqn. (1) for FER into a vector form: $r_0$ becomes $F_i(r)$, a probabilistic point forecast of rainfall $r$ from member $i$ for the period/gridbox in question, $G_i$ is that member's raw rainfall forecast, and FER becomes $M_i$(FER), the mapping function selected for that member. With rearrangement:

$$F_i(r) = (1 + M_i(\text{FER}))G_i \qquad \text{where} \quad M_i \in \{M_1, \ldots, M_m\} \qquad (3)$$

The final probabilistic rainfall forecast vector $F$, for a point in a given gridbox, is then simply derived using all n ensemble members:

$$F = \frac{1}{n}\left(\sum_{i=1}^{n} F_i(r)\right)$$

The ensemble of ensembles computed operationally currently utilises m=214 mapping functions, with the FER pdf for each simplified into 100 possible outcomes, for computational speed. So for each gridbox/period we arrive at 5100 possible realisations (100 * 51 members), which are then distilled into percentiles 1,2...99 for forecasting.

## Verification Results

To have value PP techniques must improve upon raw NWP forecasts. So the performance of 1 year of retrospective forecasts from both systems was compared, using as truth 12h rainfall observations from both standard SYNOP reports (global coverage) and specialised high-density datasets (certain countries, mainly European[34]). Verification and calibration periods were separate. Although raw model output does not pertain to point values, it is very common to verify in this way, as in ECMWF's two headline measures for precipitation[35,36].

Here we utilise categorical verification because threshold setting is common for applications and because forecast products reflect this (Section 5 and Methods). In this framework the two fundamental aspects to assess are reliability (i.e. under or over-forecasting) and the capacity to discriminate events. For these we use respectively the Reliability component of the Brier Score[37] and ROCA (Area under the Relative Operating Characteristic curve[38]). When verifying probabilistic grid-based forecasts against point measurements once cannot achieve perfect discrimination, because of sub-grid variability. We believe that relative to other discrimination metrics ROCA, also used operationally[36], is more immune to limitations placed by this. ROCA can exhibit false skill when site climatologies differ[39], so whilst the objective here was to ascertain ecPoint's added value, by *comparing* ROCA scores, a "zero-skill" baseline, based on (local) climatological probabilities, is also needed (even if these are not available everywhere). Figure 3 displays results for three 12h accumulation thresholds: 0.2mm ("dry or not"), 10mm ("wet"), and 50mm ("extreme, with flash flood potential").

For both metrics and almost all lead times ecPoint out-performs the raw ensemble. Reliability improvements are particularly striking for the 0.2mm threshold; this relates to the dark green bars on Figs. 2a-e (i.e. $G_i*0$). The ecPoint lead-time gains[40] from ROCA, for 0, 10 and 50mm thresholds are, respectively, about 1, 2 and 8 days (centring on day 5). The particularly large gains for high totals relate, primarily, to weather-type-dependant inclusion of large multipliers for $G_i$ (see e.g. red bars on Figs. 2a-e). To put these results into context, NWP improvements have historically delivered a lead-time gain of about 1 day per decade[1,41]. To have more than "weak potential predictive strength", ROCA must be some way above the zero-skill baseline[42]. So, for 0.2 and 10 mm/12 h thresholds raw ensemble and ecPoint have potential predictive strength out to ~days 7-10. For 50mm this limit is at least day 5 for ecPoint, whereas the raw ensemble has limited utility even on day 1. ROC curves show that for 10 and 50mm thresholds ecPoint's added value, expressed as ROC area impact, stems from a better handling of the wet tail - e.g. 98th/ 99th percentiles (see also supplementary material).

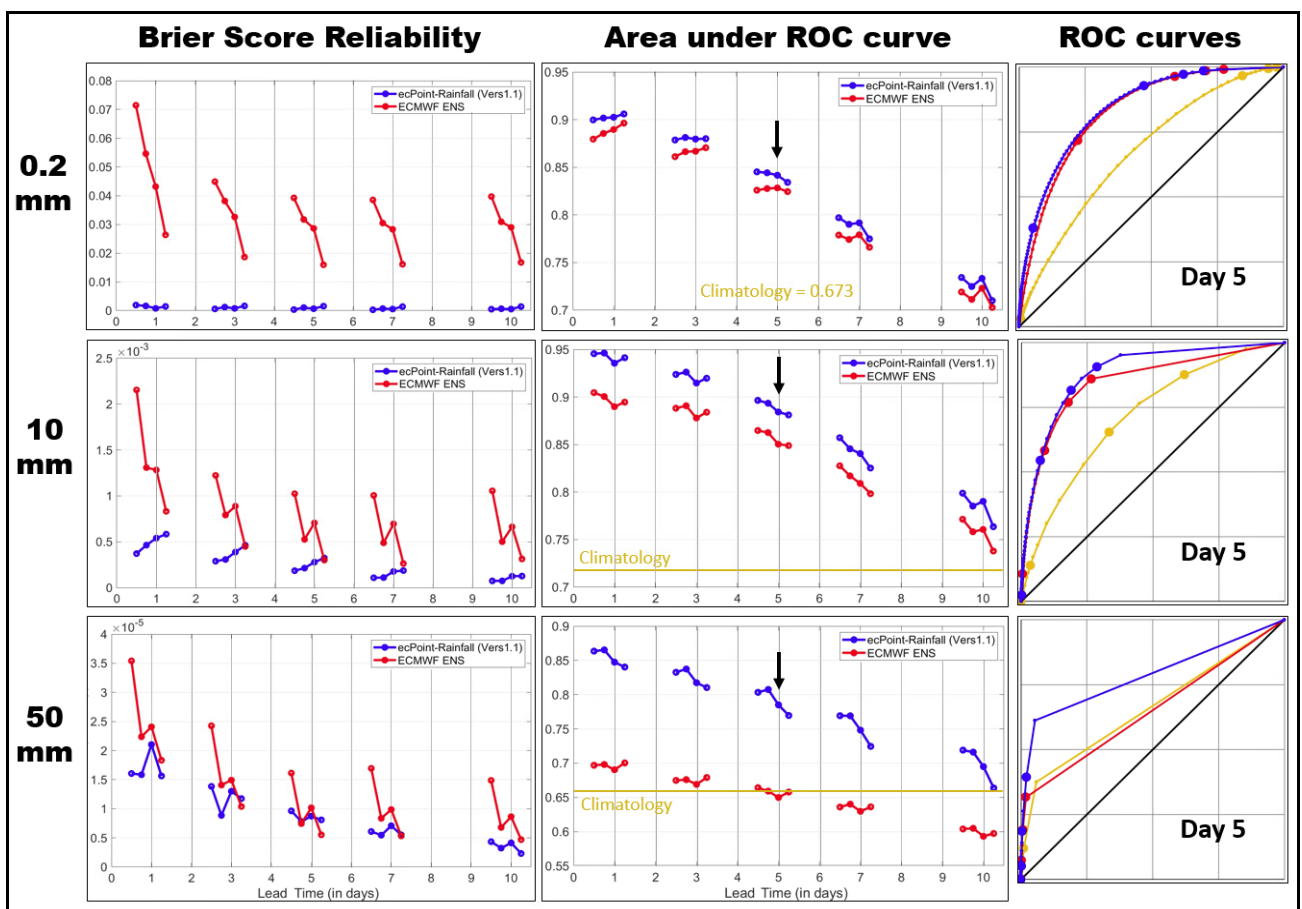

**Fig. 3**: **Category-based verification for 1 year of global gauge observations of 12 h rainfall.** Red/blue are for raw/point rainfall respectively. Rows signify thresholds of ≥0.2, ≥10, ≥50 mm/12h top to bottom. Left column shows Brier Score reliability component, 0 is optimal; points intersecting vertical gridlines denote 12 h periods ending 00 UTC, others are for end times (left to right) of 12, 18, 06 UTC. Central column shows area under the ROC curve as a measure of discrimination ability, larger is better, upper limit is 1; point meaning as for left column; yellow denotes a "baseline", for climatology-based forecasts (for sites where that is available); arrows denote lead time used for the right column. Right column shows ROC curves for day 5 (false alarm rate=0 to 1 (x) versus hit rate=0 to 1 (y)), including climatology; large spots signify probabilities of 2 (topmost), 4, 10 and 51%. Percentiles 1, 2…99 are used for point rainfall and for climatology.

Equivalent ROCA plots for the tropics only (not shown), where site climatologies should be more similar, indicate for point rainfall larger absolute increases in ROCA and larger lead time gains. Nonetheless similar plots for just extratropical regions (not shown) exhibit lead time gains only ~25% less than the quoted global values. Overall the slightly better tropical performance arises because for convective weather types, which are more common here, there is more value to be added, which ecPoint does.

Reliability, particularly for the raw ensemble, is better for leads > day 1, which may relate to ensemble perturbations being optimised for the medium range. A model spin-up issue also affects the leftmost points (T+0 - 12h). The imperfect ecPoint reliability for large totals probably relates to information loss in the distribution tail, arising because the largest percentile verified is 99$^{th}$, even though 99.98$^{th}$ is computed. Although areal warnings of flash flood risk are now issued for relatively low point probabilities we nonetheless expected little user interest in chances of < 1 in 100, and products and verification reflect this. Indeed, ROCA values for ecPoint (and climatology) could both be improved, at low cost, by including higher percentiles, but in practice impact on users would probably be small.

The oscillations seen on all panels on Fig. 3 are a function of UTC time, and relate to irregular observation density coupled with outstanding systematic errors in forecasts of the diurnal cycle of convection[31,43]. Ongoing work with a 6 h accumulation period and a governing variable of local solar time should help ecPoint to address this deficiency.

## Case Study Examples

On 25$^{th}$ February 2019 cyclonic weather (Fig. 4e) delivered extreme rainfall to parts of western Crete: >75mm/24h was widely reported, with up to 373mm/24h locally (Fig. 4f). Extensive damage occurred, including the collapse of 2 bridges[44]. In the preceding days raw ensemble rainfall forecasts (Fig. 4a) were noisy, jumpy and confusing, with relatively low probabilities of large totals. In relative terms ecPoint (which of course uses the raw ensemble) performs much better (Fig. 4b). First, fields are smoother. Second, probabilities grow over time, are more consistent and are less jumpy. Third, ecPoint's largest probabilities were more focussed on western Crete where floods occurred. And fourth, probabilities are higher overall. Conversely, we expect an orographic enhancement shortfall in the raw model (due to much lower mountains, Fig. 4f) which was apparently not rectified by ecPoint. Nonetheless, Figs. 4a,b suggest that forecasters could have provided much better warnings here using ecPoint fields, and verification results on Fig. 3 for 50mm/12h suggest that this conclusion has general validity.

The probabilities for a large threshold are usually higher in ecPoint output because PP tends to extend the wet tail, at least in (partially) convective situations, as on the cumulative distribution function (CDF) in Fig. 4c. This extension requires, for mathematical reasons, some intra-ensemble consistency. In an alternative case of wet outlier(s), probabilities within the tail can be reduced - e.g. Fig. 4d. This is because gauge totals following convection are most likely to be *less* than the forecast value - e.g. see Fig. 2.

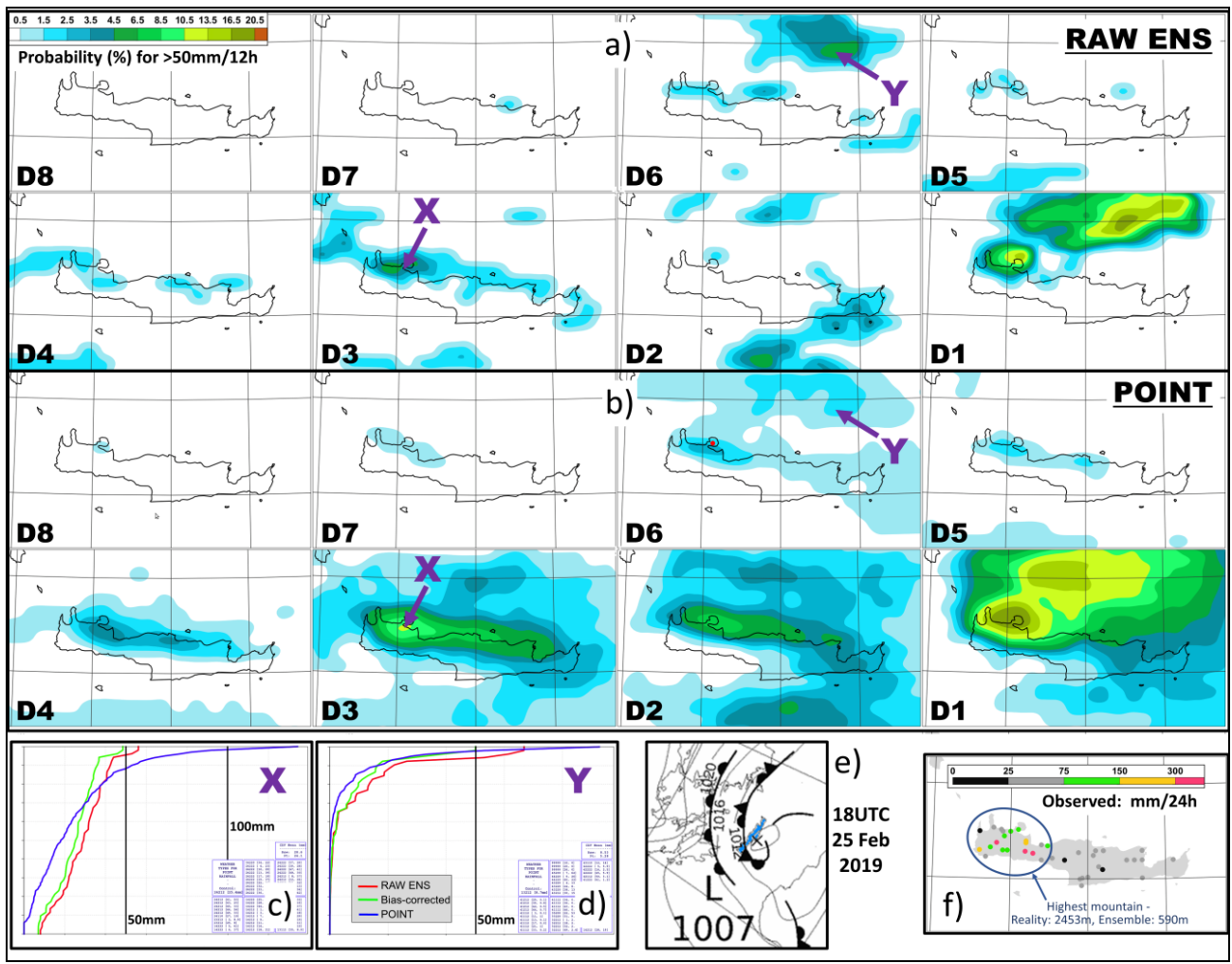

**Fig. 4**: **Evolution of raw ENS and ecPoint forecasts for a flooding event in Crete**. **a)** Forecast probabilities (%), for 12-24 UTC 25th February 2019, for rainfall >50mm, from raw ensemble; D8, D7, …D1 (for days 8, 7, …1) denote data times of 00UTC on respectively 18th, 19th … 25th February. **b)** As **a** but for ecPoint. **c)** CDFs for 12h rainfall from the respective D3 forecasts for site X shown on **a** and **b**, for raw ensemble gridbox totals (red), bias-corrected gridbox totals (green) and point rainfall (blue), for 12-24UTC 25th (*y*-axis spans 0-100% and lies at *x*=0mm). **d)** as **c** but for D6 for site Y. **e)** UK Met Office surface analysis for 18UTC 25th (Crete in blue). f) Gauge observations of *24h rainfall*: 00-24UTC 25th February 2019 (hardly any 12h totals are available in the area).

Another feature of ecPoint PP is illustrated on Figs 5a,b which are for a cyclonic, convective, winter-time case in Norway. Typically, ecPoint's wetter tail will cross the raw ensemble tail around 85 % (e.g. Fig 4c). But here, even for the 95th percentile, in the north-south chain of maxima (Fig. 5a), ecPoint values are still lower. The reason here is not outliers; it is instead the adjustment for an expected large over-forecasting bias: compare green and red lines on Fig. 5b. The main weather type is "44210" - note how its C value of 0.47 is particularly low (Figs. 2d,f). Norwegian forecasters, based on experience, also envisaged that rain in this situation would be over-forecast (personal communication, Vibeke Thyness). Nearby gauge observations (black dots, Fig. 5b), and radar-derived totals (not shown) seem to support better the ecPoint forecasts.

Fig. 5c shows that training data for type "44210" came mostly from mountainous regions near coasts, consistent with topographic convective triggering being the rainfall generation mechanism, and with this being overdone in the model. This seems to be because when low-CAPE airmasses impinge on mountains convective cells take time to grow and rain out, whilst in global NWP rainfall is immediate. ecPoint highlights quantitatively the impact. In turn the large biases identified raise important questions about how, for example, related latent heat release overestimates might reduce subsequent broadscale predictability. The Crete case (Fig. 4) differs because rainfall there had a smaller convective component.

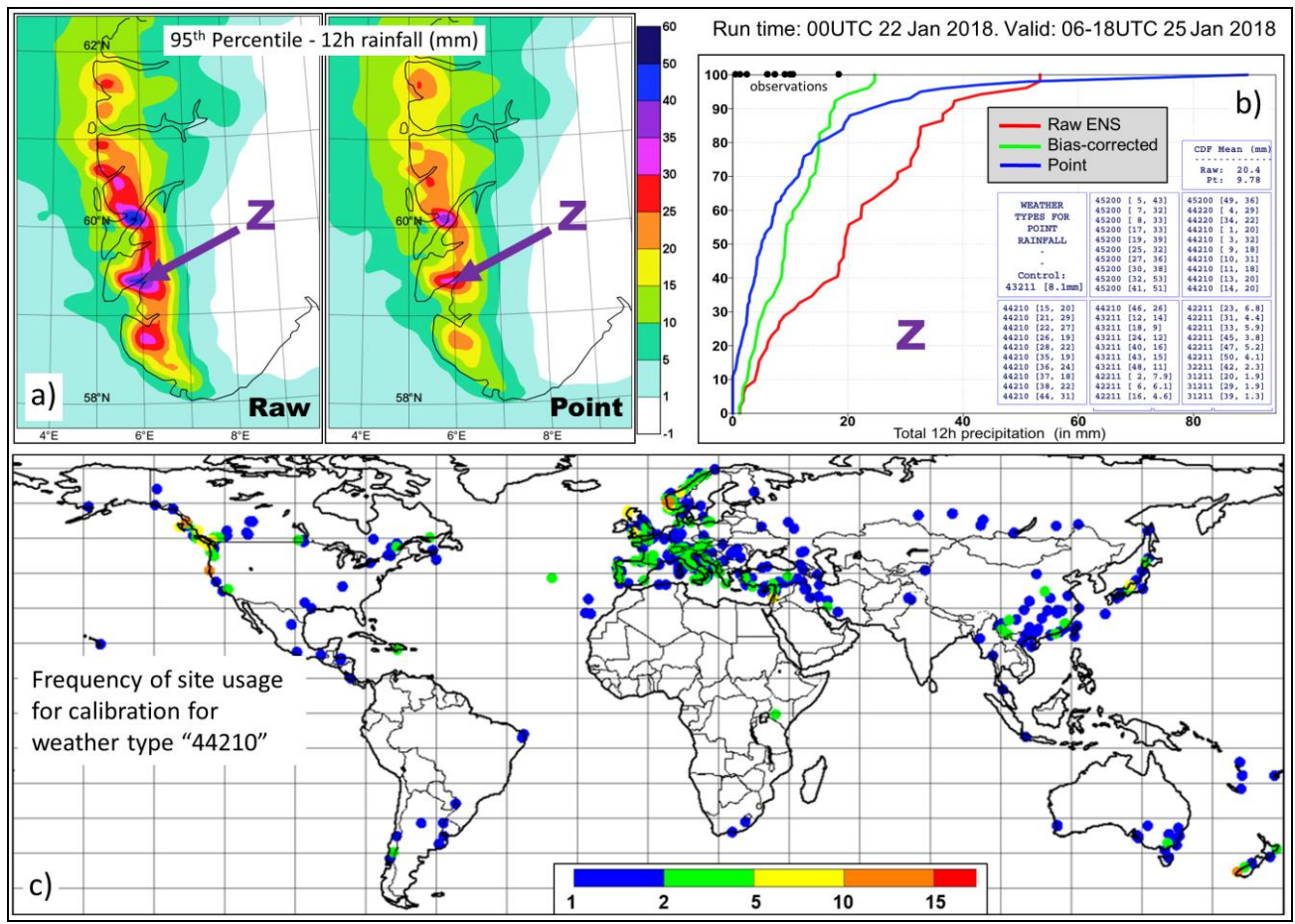

**Fig. 5**: **a) Day 4 rainfall forecasts for southwest Norway, with a calibration site usage map. a)** 95th percentile of 12 h rainfall (mm) from data time 00UTC 22nd January 2018 for 06-18 UTC on 25th, from raw ensemble (left) and ecPoint (right). **b)** CDFs of 12 h rainfall for site Z shown on **a**; same data/valid times as **a**; colours as Figs. 4c,d; table inset shows member number, forecast rainfall in mm, assigned weather type as a 5-digit code. **c)** Locations of calibration events used to define the mapping function for code "44210" (see Fig. 2d).

These examples show how ecPoint is rich in physically realistic complexity, which helps deliver much better forecasts on average. Similarly using gridbox-weather-types increases understanding of model behaviour. Other PP methods do not generally have these desirable characteristics.

Figures 4a,b and 5a (and Extended Data Figs. 2a and 3a) provide examples of experimental ECMWF products being delivered to forecasters since April 2019. Users can select their own thresholds, compare raw and point forecast fields, and reference user guidelines[45].

## Discussion and summary

We describe a completely new PP approach to forecasting weather at sites - ecPoint - to meet customer need in ways that pure global NWP cannot. Our focus has been on rainfall (renowned to be particularly challenging[46,47]) but the philosophy applies also to other variables. Verification results are very positive, and new product characteristics will be appealing to users. Uniquely amongst PP methods, ecPoint output covers the world.

One methodological aspect has been pivotal: the use of gridbox-weather-types in lieu of location-specific PP. This delivers, relative to other studies, vast calibration datasets that retain physical meaning; the many other attractive features of ecPoint stem from this (Table 1). "Weather type" importance has long been recognised, but in the form of country-scale circulation patterns[48,49] which imposes huge constraints on training data size compared to our gridbox-weather-type approach.

Whilst ecPoint can predict local extremes via "remote learning", the lack of *global record* conditions in training data could very occasionally be a constraint. However, using mapping functions means that new global records *can* be predicted whilst in some other PP methods they cannot[12].

For *specific sites* alternative MOS-type techniques might provide better forecasts than ecPoint, if local topography and/or meteorological characteristics are especially unusual. Similarly, local verification will inevitably reveal some regional differences in ecPoint performance. These should however stimulate future upgrades to the governing variables and the decision tree. More complexity can certainly be included, but in time greater observation coverage will be needed to support this. We therefore plan to exploit more high-density observations, including new crowdsourced data[50].

Whilst the simple, expert-driven creation of our decision tree has proved very powerful, there is clearly scope for optimisation. Carefully tailored machine learning tools[51] could be exploited, alongside more complex decision tree branching, and may even become a necessity if governing variable count increases. However, we caution against black-box methods, and enforced statistical complexity, for fear of eroding the capacity to deliver meaningful physical insights. It is vital for NWP development to understand the differences between raw and bias-corrected gridbox values. Figures 5b,c provided a clear example of why. Such insights are a major advantage of the ecPoint approach.

ecPoint has many other diverse applications. Its output is currently being blended with post-processed high-resolution limited-area ensemble forecasts[52], to exploit their better representation of some topographic effects (reference Fig. 4f), to deliver better forecasts overall and to seamlessly transition into the medium range. Meanwhile historical probabilistic point rainfall and point temperature re-analyses will be created from the gridscale global re-analysis called ERA5[53], delivering unparalleled representativeness back to 1950. The pointwise climatologies so-derived will themselves have numerous applications, such as providing a reference point (independent of model version/resolution) to see if ecPoint forecasts are *locally* extreme, via the "extreme forecast index" philosophy[54]. Thereby ecPoint forecasts become even more relevant for flash flood prediction. Other applications include inexpensive downscaling of climate projections[55], global tests of hypotheses such as "do cities affect rainfall"[56], and quality-control of observations[57] using a point-CDF-based acceptance window. Improved bias-corrected feeds into hydro-power and hydrological[58] models are also objectives.

So why is ecPoint "low cost"? Providing control run forecasts for calibration is orders of magnitude cheaper than providing the multi-year ensemble re-forecasts needed by many other PP methods. And in operations, ecPoint's own computations are many orders of magnitude cheaper than global NWP alternatives (which in 2020 are very far from being operationally viable anyway). Nevertheless, even low-cost PP methods rely on NWP to provide the input. Indeed, the investment and development needs of pure NWP remain as strong as ever. The difference now is that by simultaneously exploiting the strong synergistic relationship with ecPoint we can secure much larger forecast improvements for everyone.

## Supplementary Material: Methods

Realisation of the new ecPoint prediction system required the following steps:

Calibration:
1) Use physical reasoning to decide which governing variables to test
2) Formulate the calibration assumptions
3) Create a calibration dataset
4) Test the utility of the governing variables using the calibration dataset
5) Select breakpoints for each governing variable, and create a decision tree
6) Construct one "mapping function" for all variable range combinations

Forecast production:
7) Create an operational ecPoint system using mapping functions
8) Devise output formats that can be effectively used and verified

### **Physical reasoning**

Simple physical reasoning, meteorological experience and case studies can immediately suggest variables that are likely to influence sub-grid variability and/or model bias, and which therefore each have potential to become a governing variable used to define the gridbox-weather-types. In putting together such a list, for testing, there are two additional considerations. Firstly, avoid double counting - i.e. different variables addressing the same model characteristic. Secondly, do not include variables that relate strongly to rainfall where the model already captures that relationship. For example, as *specific humidity* increases higher rainfall totals become possible, yet such dependence is already built into the model physics. To illustrate the potential scope and power of our approach examples of relevant variables are shown in Extended Data Table 1, in four classes. All are continuous scalar variables; some are strictly bounded, some not.

On Extended Data Table 1 parameters in class (1) clearly vary in time and are hence labelled "dynamic". The relevance of parameters 1a and 1b is discussed in the "Concept" section in the main text. Convective available potential energy (1c) can control the vigour of convective cells, which can potentially influence bias or sub-grid variability, whilst total precipitation (1d) is included to address any limitations of expressing FER in ratio form (Eqn. 1). Higher vertical wind shear (1e) can favour longer-lived convective cells[59] and as this aspect is not modelled it can potentially cause model error. Vertical motion computed by inverting the omega equation[60–62] (1f) brings the opportunity to disentangle dynamically-forced from orographically-forced rainfall, the latter being more prone to model-resolution-dependant errors. In case studies positive values (for 1f) have also increased the longevity, relative to model representation, of convective systems once diurnal heating diminishes. Convective adjustment timescales (1g) have been shown to relate to predictive skill in convective regimes[63–65], which could prove useful for refining sub-grid variability. Meanwhile Corfidi vectors[66] (1h) might provide an alternative, improved estimate, relative to 700hPa wind speed (1b), of the convective cell movement speed, with implications for sub-grid variability as discussed in the "Concept" section.

Class (2) includes topographic sub-grid complexity (2a) because this can affect the distribution of rainfall at the sub-grid level. Meanwhile population density (2b), not currently represented in the ECMWF IFS (integrated Forecast System), may create a heat island effect that influences convective precipitation overhead and nearby[56]. These variables are clearly "static" rather than dynamic (during a forecast).

The astronomical class (3) is important because insolation strongly influences the development and decay of land-based convection. We consider insolation parameters (3a,3b) *as well as* convective fraction (1a) in part to distinguish convective precipitation that cannot have been triggered by insolation, because that might have different sub-grid variability and model bias characteristics associated. In respect of local solar time (3b), there are well-known biases in the diurnal cycle of convection[31,43], implying different FER characteristics at different times of day.

Class (4) is model dependant and is more advanced in that different variables are combined in ways steered by meteorological understanding. For example, we know that orographic influence on large-scale rainfall[67] (4a) can depend strongly on slope-normal wind strength, so a governing variable representing this can bring into play sub-grid variability, and biases linked to orographic enhancement and rain shadow . One would expect a priori that all three would increase with the value of this variable. A cell-drift parameter (4b) is listed to address the "zero lifetime" weakness implicit in convective parametrisation: convection cannot move beyond its triggering source region and so convective rainfall will often stop, unrealistically, at coasts[45]. Accordingly, one can invent gradient-related governing variables, such as the dot product of a steering wind and the gradient of the land-sea mask, to isolate out regions vulnerable to such errors. Such a variable can in effect spread model information beyond gridbox edges in realistic ways.

### Calibration assumptions

In general calibration procedures used within PP compare forecast parameters with observations and rely on several assumptions. Our method is similar in these respects. Its key assumptions are as follows:

   i. Forecast biases are not a function of forecast lead time.
   ii. Random errors in unperturbed (Control) forecasts are sufficiently small at short lead times to be disregarded. In other words, such short-range predictions exhibit a relationship with observations that is consistent for a given gridbox-weather-type.
   iii. The relationships between forecasts and observations are independent of geographical location (excepting the case of indirect representation within parameter classes 2, 3 and 4 in Extended Data Table 1).
   iv. Available observations are adequately sampling true sub-grid variability and true bias for the different gridbox-weather-types.

Considering (i), in global NWP nowadays biases do not generally vary that much in the first week or so of a forecast, as shown by lead-time-based daily quantiles for the model climatology, or the "M-Climate"[45]. However, due to assimilation issues the IFS has historically had higher tropical rainfall extremes in the first few hours[68], and whilst this situation has improved somewhat over time (Richard Forbes, personal communication) we took the precautionary measure of not using the first 3 forecast hours. In future this limit is something to revisit.

For assumption (ii) to be valid we need an accurate model, with responsive assimilation, and to confine calibration to short lead times. Using the IFS Control run at leads ≥3h satisfies these requirements reasonably well.

Assumption (iii) is a re-iteration of the universality of physical laws, and recognises that latitude and longitude should not, of themselves, determine any forecast-observation relationships. This is a key difference between our method and the standard MOS approach[4]; in MOS location is everything. Whilst some recent studies have successfully relaxed this constraint via "data pooling", wherein observations from sites that are geographically close and that have for example a similar topographical aspect are grouped together for calibration[33] our method goes much further by pooling in a highly dynamic way according to governing variable ranges, without geographical constraints being imposed from the outset. In practice, the relationship between a model forecast and an observation at a site in (say) Australia in the calibration period can inform a later forecast for a site in (say) the USA, if the governing variables in model output have similar values in the two instances. This is an extremely powerful innovation because it allows the size of the calibration datasets to be increased by several orders of magnitude, compared to classical PP, and indeed is key to being able to generate many complex yet complete mapping functions using as little as 1 year of calibration data.

Assumption (iv) requires that observations are fairly evenly scattered across terrain. Mostly this will be true, although in mountainous regions, for example, there may overall be more data per unit area for lower elevations.

Whilst it is difficult to test in a comprehensive and direct way the true validity of the above four assumptions the very striking verification results, detailed in the "Verification" section of the main text, strongly suggest that these assumptions were a reasonable basis for our calibration.

### Calibration dataset

Here a "calibration dataset" serves two purposes; it allows for potential governing variables to be tested, and it subsequently defines the mapping functions decided upon. It must contain rainfall observations and governing variable values.

Firstly, we accrue rain gauge observations from around the world, for the rainfall period in question (e.g. 6 or 12 or 24 h), typically spanning a 1-year period. These comprise standard observations available through international data-sharing agreements, plus any others available from special datasets[31,34]. Higher volumes of observational data will ultimately deliver better results. Secondly, to pair up, one needs short range control forecast output from the current model version, naturally spanning the same period as the observations, and incorporating values of forecast rainfall and all governing variables under test for all observation dates. Creating this for each new model version is straightforward, and is computationally cheaper, by about 3-4 orders of magnitude, than creating 10 years or more of ensemble re-forecasts as used in some other PP techniques[9,15,18,27,58,69–71]. Re-forecast utilisation has previously been advertised as *the way* to improve statistical forecasts[72].

The calibration dataset is then a large tabular file in which each row represents one observation for one site. Columns are "attributes", comprising observed rainfall, rainfall forecast by the closest valid Control run, values of all the candidate governing variables in that Control run (computed as a time average or an accumulation or some other function) and finally the FER value itself (computed using Eqn. 1). Forecast rainfall is the value at the gridpoint nearest to the observations (i.e. *not* interpolated). Other governing variables may be interpolated or not according to meteorological considerations.

To minimise any detrimental effect on FER distributions caused by observation discretization (typically 0.1 or 0.2 mm for rain gauges) we then remove all instances when forecast rainfall is <1 mm.

### Test the utility of governing variables

The null hypothesis under test here is: "neither sub-grid variability nor bias, as represented by the FER distributions, *depend* on the governing variables". In principal there are many ways to test using the tabulated data, from fully automated machine learning[51] through to a wholly subjective approach. We adopt a straightforward, transparent and semi-subjective methodology, examining candidate variables individually. For each we use dual-subsets separated by an escalating breakpoint[1]† (starting at the dataset minimum). At each such value we first check for adequate sample sizes, and then for the two FER distributions examine the following numerical metrics to see if those distributions differ in a substantive way:

i. Two-sample Kolmogorov-Smirnov test
ii. Relative frequency of "mostly dry" (i.e. using FER < -0.99)   [dark green on Fig. 2]
iii. Relative frequency of "over-prediction" (i.e. using FER < -0.25)   [both greens on Fig. 2]
iv. Relative frequency of "good forecasts" (i.e. using -0.25 < FER < 0.25)   [white on Fig. 2]
v. Relative frequency of "substantial under-prediction" (i.e. using FER>2)   [red on Fig. 2]

The inclusion of items (ii) to (v) list helps address known weaknesses in the Kolmogorov-Smirnov test, such as being insufficiently responsive to differences in the tails. Compared to using an alternative statistic to address this, such as the two-sided Anderson-Darling test[73,74], our simple strategy usefully provides clear sight of which aspects differ in the subsets.

The tests revealed a very clear dependence of FER distribution on 5 out of the 9 governing variables starred on Extended Data Table 1. These were CF, TP, $V_{700}$, CAPE and $S_{24}$. The other 4 variables, for which the null hypothesis could not be easily rejected, were set aside, pending possible re-formulation and re-examination at some future point.

### Create the decision tree

The next objective was to divide up all FER values into sets (i.e. distributions), using each of the 5 selected governing variables, and by delineating using user-defined breakpoints. This must lead to one distinct FER distribution, i.e. mapping function, being defined for all possible governing variable range combinations. Whilst acknowledging that there are many possible approaches[51], we elected to construct a straightforward and easy-to-comprehend single decision tree, in which each of the 5 branch levels corresponded to one variable. Steered by the results from (d) above, by the original ecPoint concept, and by analysis of multiple case studies, we selected from the outset the following level hierarchy (top to bottom, most significant at the top): CF -> TP -> $V_{700}$ -> CAPE -> $S_{24}$. Extended Data Fig. 1 shows a small portion of the final decision tree.

---

† Also known as "cutting values" in decision-tree literature.

The decision tree was constructed, in systematic fashion, as follows. First decide upon the breakpoints for the first variable and accordingly create branches to level 2. Then consider in turn each of those branches. For each of these decide upon breakpoints for the second variable, and thereby create sub-branches to level 3. And continue using the same approach for the third, fourth and fifth variables.

The key decision in the above is choosing the breakpoints. Each such value is selected for one of two reasons: either it denotes a discontinuity in FER distribution characteristics, or in the case that those characteristics appear to be evolving in more continuous fashion (which happened less often) it denotes a convenient point, from the perspective of subset sizes, at which to divide. We also tried to retain breakpoints that looked numerically sensible, recognising that high precision is hard to justify (e.g. "2" might be chosen instead of "2.1").

The approach and the metrics used for breakpoint selection were as listed in section (d) above. In practice the procedure involved some compromises, that can be usefully illustrated via Extended Data Fig. 1. At level 1 the two-subset analysis provided justification for three breakpoints (rather than some other number), and these were close to 0.25, 0.5 and 0.75. The same procedure was then used for TP at level 2, for each of the four CF categories. Within the subset of FER cases with CF>0.75 (highlighted) there were noteworthy FER distribution differences right across the range of TP values, but due to diminishing sample size for larger totals a compromise was made in creating the wetter categories such as TP>32 mm. It was clear that for those it would then not be possible to sub-divide much further at levels 3 to 5. In the end, whilst the set (CF>0.75, 2<TP<8) lead to 16 branches at level 5 (i.e. 16 mapping functions), the set (CF>0.75, TP>32) lead to only 3, with no dependence at levels 4 and 5. This is not necessarily because the case (CF>0.75, TP>32) had no real dependence on CAPE and $S_{24}$; rather the sample was too small to test this.

### Construct mapping functions

The calibration procedure output comprises the decision tree, and one mapping function for each leaf. Evidently each mapping function is a distribution of FER values (Eqn. 1) that themselves each represent one rain gauge measurement taken somewhere in the world at some time during the calibration period, when the gridbox forecast rainfall was ≥1mm. To be used for a given mapping function a measurement must correspond to a time when the governing variable range inequalities, defining the mapping function in question, were all satisfied in the corresponding short-range Control forecast for the requisite gridbox.

An important constraint when creating mapping functions is that they should not be particularly prone to sampling issues, and this was achieved by retaining relatively large samples (≳200) at leaf level in the decision tree. It is clearly beneficial to have many observations in the calibration period, as for all PP methods[17], and as ~1.7 million FER values were created for calibration the 214 mapping functions denote, on average, 8,000 cases. So mapping function case count varies greatly. The fact that the most-used mapping functions will clearly have the highest case counts is a robust and beneficial feature of our approach.

So how do we store a mapping function - as visualised on Figs. 2a-e - to make it usable by our PP procedure? We first divide up its numerically sorted list of FER values into j almost-equally-populated subsets, and then select a representative FER value for each. The value j was assigned from the outset, through the following considerations:

i. j be large enough for the full distribution to be well represented, including the wet tail,
ii. subset size be large enough to minimise sampling noise, and
iii. j be small enough to facilitate timely forecast production.

In practice item (iii) proved to be the main constraint and following some experimentation and consideration of supercomputer resource we set j=100, for all mapping functions. All representative FER values (i.e. 100 * 214 mapping functions) are stored in one calibration matrix (along with the inequality definitions) to use whenever a real-time ecPoint forecast is computed.

### Operational forecasting system

Equipped with the calibration matrix, and the predicted rainfall and governing variable values as input from each ensemble member (51 in our case), one can compute the ecPoint rainfall distribution forecasts for every gridbox in the world. For one gridbox, in one ensemble member, for one time interval, these are the steps:

i. compare the forecast input data with inequality thresholds in the calibration matrix to decide which one of the 214 gridbox-weather-types characterises that gridbox
ii. using the raw rainfall forecast and the 100 representative FER values for the said weather type together in equation (3) compute 100 equiprobable point rainfall values

We repeat (i) and (ii) here for each ensemble member (including the Control run) and then combine and sort in ascending order the resulting 5100 numerical values. This could be the final gridbox output but because of storage constraints and usage considerations we divide into 100 equally-populated subsets, and save instead the percentiles 1, 2,..99, where percentile 1, for example, bisects values 51 and 52 in the sorted list.

When considering all timesteps, full global coverage, and an operational necessity to complete in under 1 hour, the PP procedure becomes computationally challenging. The main memory hurdle to overcome is retention of 5100 forecasts at one time, whilst the main computational challenges include assigning the gridbox-weather-types (timings being proportional to the number thereof), sorting, and limiting read-write operations and temporary files. Fortunately, the computations lend themselves to simple parallelisation on a supercomputer platform. In practice we choose to divide the world into 10 blocks at the memory intensive stage, before re-combining at the end. We expend a total of about 21 CPU hours to complete for one set of forecast runs, which parallelisation reduces to <1 hour run time.

For developers with different computational resources many options exist for altering memory consumption and clock time, such as changing the number of weather types and/or the number of representative FER values. There are naturally disadvantages of reduced computer resources, which would be seen in verification metrics. Conversely increased resources bring advantages, although these are ultimately capped because the number of observations available for calibration imposes its own constraint. In these respects, PP behaviour broadly mirrors that of NWP.

### Output formats

ECMWF purveys its forecast data to national meteorological services (NMS), which includes delivering map-format charts to forecasters. Formal release of ecPoint rainfall output has been through this channel, on the ecCharts platform[45], although experimental provision of gridded data to a few NMS preceded this[75].

A primary design consideration for new products is what mode of use could benefit society most. Here that is when ecPoint performs much better. Two opposite instances arise, as highlighted by verification (Fig. 3): localised rainfall extremes, that may cause flash floods, and mostly dry weather, for activity planning for example. These needs are satisfied by facilitating display as either a user-selected percentile in mm (from the 99 available), or as a computed probability in % for a user-selected threshold (bi-directional, in mm).

Distribution tails are ordinarily much more pronounced in ecPoint rainfall. Clear depiction of these is paramount, to highlight but not over-emphasise extremes. For each format we therefore created a non-linear numerical scale with tapered colour saturation representing the compressed-interval tail. The scheme was otherwise polychromatic, to facilitate rapid value identification throughout the range - a standard user requirement. Identical options were also introduced in ecCharts for the raw ensemble, for comparison, and as a learning tool.

Extended Data Figs. 2 and 3 provide examples from May 2019 of operational ecCharts-style output (a), illustrating also how forecasts for a given date can evolve as lead time shortens, and allowing comparison with observations (b, c). Extra-tropical (Fig. 2) and tropical (Fig. 3) regions are depicted, to illustrate how behaviour varies when climatologies differ. Plot design allows users to immediately identify the following:

- The raw ensemble shows generally higher probabilities than ecPoint for totals >0.5mm, particularly in the tropical domain (compare 1st and 2nd columns on Extended Data Figs. 2a, 3a). The frequencies of observed totals >0.5mm (Extended Data Figs. 2b, 3b) generally match ecPoint probabilities better.
- The ecPoint 99th percentile fields (Extended Data Figs. 2a, 3a; 4th column) generally show higher values, by some margin, than the locally most extreme ensemble member (3rd column). Sometimes the difference is more than a factor of two. The most extreme observations (Extended Data Figs. 2c, 3c) tend to be better captured by ecPoint than by the raw ensemble. This is true over central and southern Germany where floods and flash floods were reported[76].
- There is more variation between successive forecasts in the extra-tropics than in the tropics. This is due to a greater dependence of rainfall, in the extra-tropics, on specific weather systems, coupled with day-to-day variations in forecasts of those systems.

These observations are all in agreement with what one expects, meteorologically, from the ecPoint system. Future options for complementing, operationally, the output formats shown on Extended Data Figs. 2a and 3a, include site-specific graphs such as the CDFs shown Figs. 4c,d and 5b. Map-based depiction of bias-corrected grid-scale forecasts is also an option.

Other types of customer-oriented product can be devised, such as "guideline rainfall in mm for the wettest point in a model gridbox", using the median of the 99th percentiles for each ensemble member (from $F_i$(r), eqn. (3), main text). This would require a computational intervention after applying eqn. (3), and a creative approach to any verification thereof. Similarly, to better address hydrological needs one could in principal perform a gridbox-location-dependant adaptation of the post-processed output for each ensemble member to reflect local catchment size. For example, if catchment size was ~25% of gridbox size one could, using physically meaningful rainfall aggregation assumptions, distil the (now 100) post-processed "point" realisations into 4 equi-probable realisations to represent total rainfall averaged over a "quarter-size gridbox" for each member.

**Data Availability**

Data used for this study (except Fig. 1) were generated by or collected at the ECMWF large-scale facility. Derived data supporting the findings of this study are available from the authors on reasonable request.

**Code Availability**

For the post-processing code, calibration files, images of mapping functions, and a link to data samples on Zenodo see: https://github.com/ecmwf/ecPoint/releases/tag/1.0.0.

# Supplementary material: Discussion

Figure 3 highlighted how ecPoint delivered striking improvements in ROCA for large totals (50mm/12h), indicating much greater "potential predictive strength"[42]. To better deconstruct this important result we depict in Extended Data Fig. 4 three ROC curves for a short (24h) lead time. At such leads we expect a more illuminating picture to emerge, because divergence within the ensemble, in rainfall totals and weather types, is relatively small then. In turn this means that within the ecPoint calculations the wet tails belonging to each ensemble member's CDF will reinforce rather than dilute each other, making it more likely that an extreme will be captured within integrated full-ensemble percentiles that we store (1..99). In the case of more divergence the converse would hold; extremes would be "predicted" more often overall but the associated probabilities would be lower, and lost in the current configuration whenever they were <1%.

Extended Data Fig. 4 shows that about half of the above-diagonal ROCA for ecPoint and the raw ensemble can be attributed to probability levels of 1,2,3,4% and 2,4,6% respectively. Meanwhile for climatology the bulk of the above-diagonal ROCA comes from just the 1% level. So, although raw ensemble and climatology have here a similar ROCA (Fig. 3) the former is associated with much higher probabilities. Then directly comparing points for equal probabilities between the raw ensemble and ecPoint, we see that ecPoint achieves more ROCA at a given probability level for all probabilities up to about 8%. By day 5 (Fig. 3) that crossover reduces to about 4%, but by then higher probabilities are contributing relatively little. In fact the ROC curve "trend", with increasing lead time, involves points for a given probability level translating towards the origin. This applies to both ecPoint and the raw ensemble, but not to climatology which is of course fixed, and which therefore becomes increasingly competitive as a "predictor" for longer leads, as expected.

For a more complete picture of the full utility of different systems one would need to account for different cost-loss ratios[30], which is beyond the scope of the present study. Nonetheless, this discussion confirms that for large thresholds the discriminating ability of ecPoint is largely associated with its wet tails. More of this capability (as measured by ROCA) could probably be retained at longer lead times by introducing probabilities below 1%, though how genuinely useful they would be in themselves, and against climatological forecasts extended in the same way, is not so clear. This is a topic for future work.

# Extended Data

**Extended Data Table 1**: Some candidate governing variables, in 4 classes.

| Class 1 = "*Dynamic Raw Model*" | | Shorthand |
|---|---|---|
| **1a*** | **"Convective precipitation fraction" [during period]** | CF |
| **1b*** | **"700 hPa wind speed" [average of values during period]** | $V_{700}$ |
| **1c*** | **"Convective available potential energy" [maximum during period]** | CAPE |
| **1d*** | **"Total Precipitation" [during period]** | TP |
| 1e* | "Vertical wind shear magnitude" [average of values during period] | |
| 1f | Quasi-geostrophic vertical motion (from attribution) | |
| 1g | Convective adjustment timescale | |
| 1h | Corfidi vector magnitude | |
| Class 2 = "*Static Geographical*" | | |
| 2a | "Topographic sub-grid complexity" [a model variable] | |
| 2b | "Population density" [an externally-sourced variable] | |
| Class 3 = "*Dynamic Astronomical*" | | |
| 3a* | "24 h incident clear sky solar radiation" [on given date] | S24 |
| 3b* | "Local solar time" | |
| Class 4 = "*Dynamic Computed*" | | |
| 4a* | An "orographic rainfall factor" | |
| 4b* | A "cell drift parameter" | |

* Variable already examined.
Bold type denotes variable used in the current ECMWF operational system, and in examples in this paper.

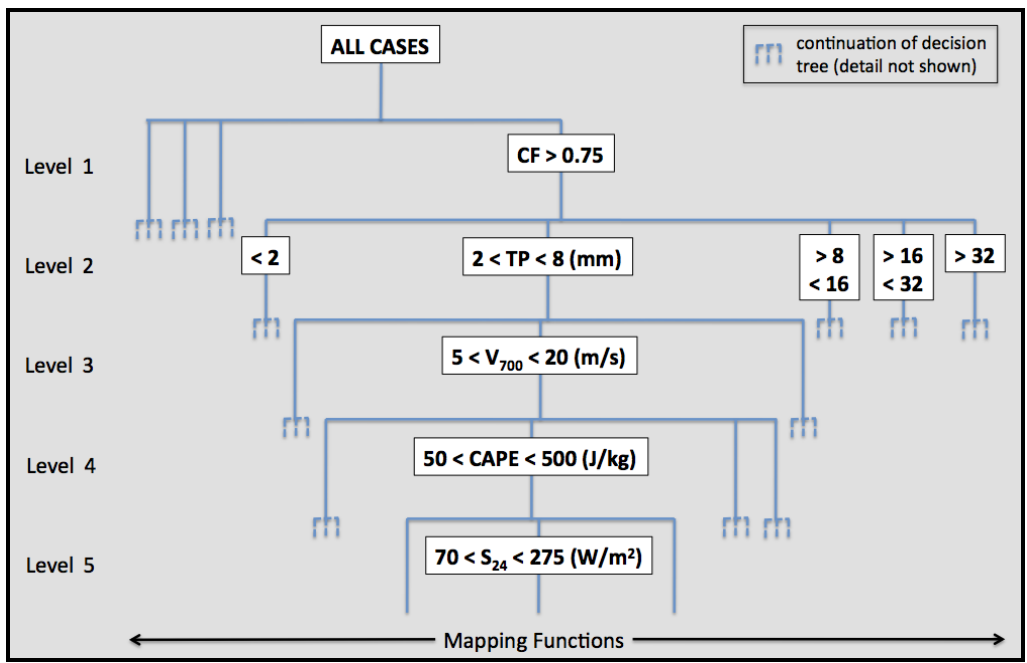

**Extended Data Fig. 1**: **A small segment of the currently operational decision tree**. Space constraints prevent depiction of every branch. In the full tree there are 214 branch terminations, known as "leaves"; most but not all are at level 5. Each leaf has one mapping function (FER distribution) associated.

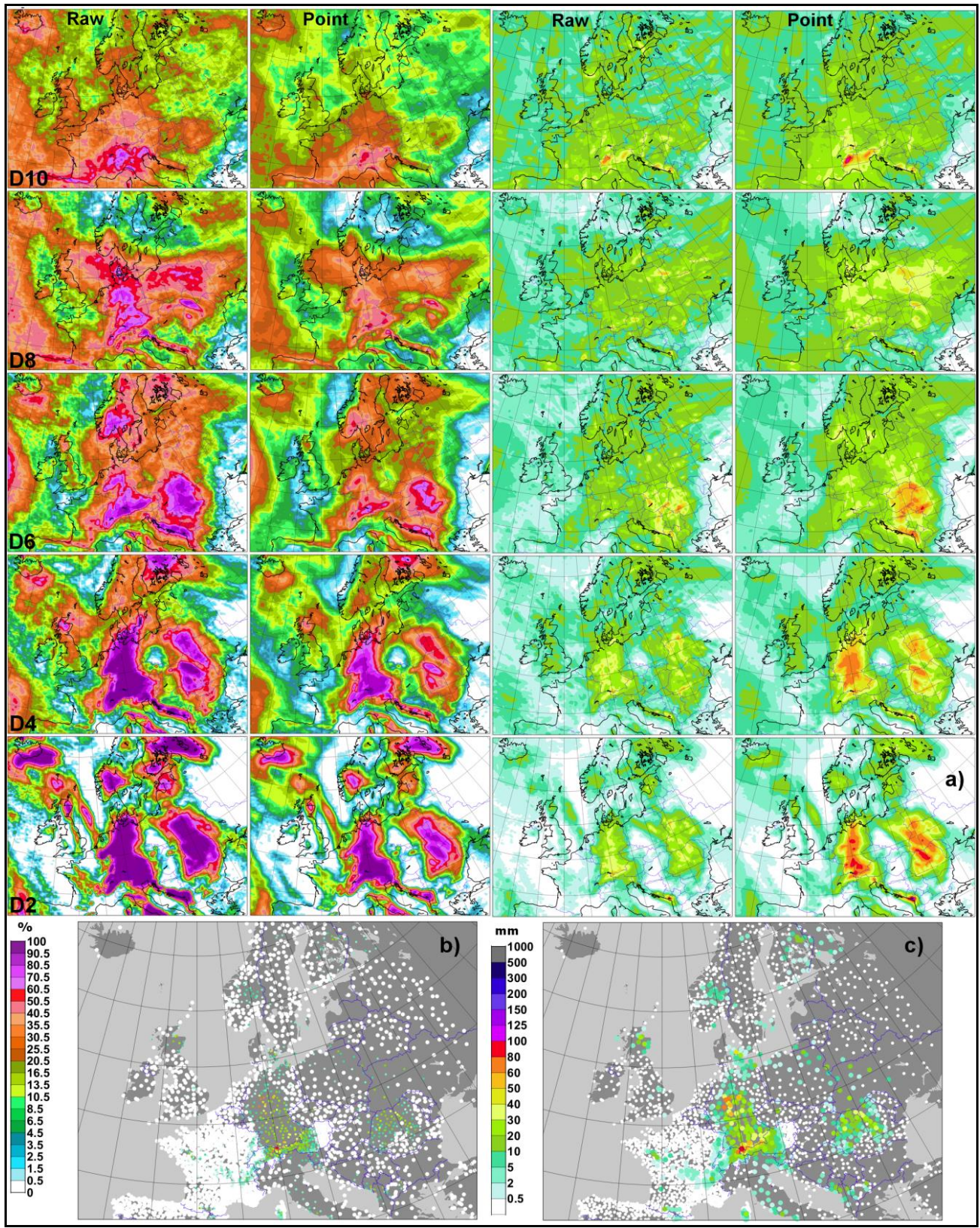

**Extended Data Fig. 2: Temporal progression of forecasts for one 12h period, for part of Europe, with verification. a)** 20 forecasts, all valid 18UTC 20th to 06UTC 21st May 2019, from start times of 12UTC on 11th, 13th, 15th, 17th, 19th May (top to bottom rows) respectively denoting lead times of ~10, 8, 6, 4, 2 days. Column 1: raw ensemble probability>0.5mm (in %, left legend). Column 2: same but from ecPoint. Column 3: maximum value in the raw ensemble (mm, right legend). Column 4: 99th percentile from ecPoint (mm, right legend). **b, c)** point rainfall observations for the same valid period as **a** (mm, right legend); plots show the same data, but spot size settings differ to address plotting constraints of a high-density network; higher values always appear on top.

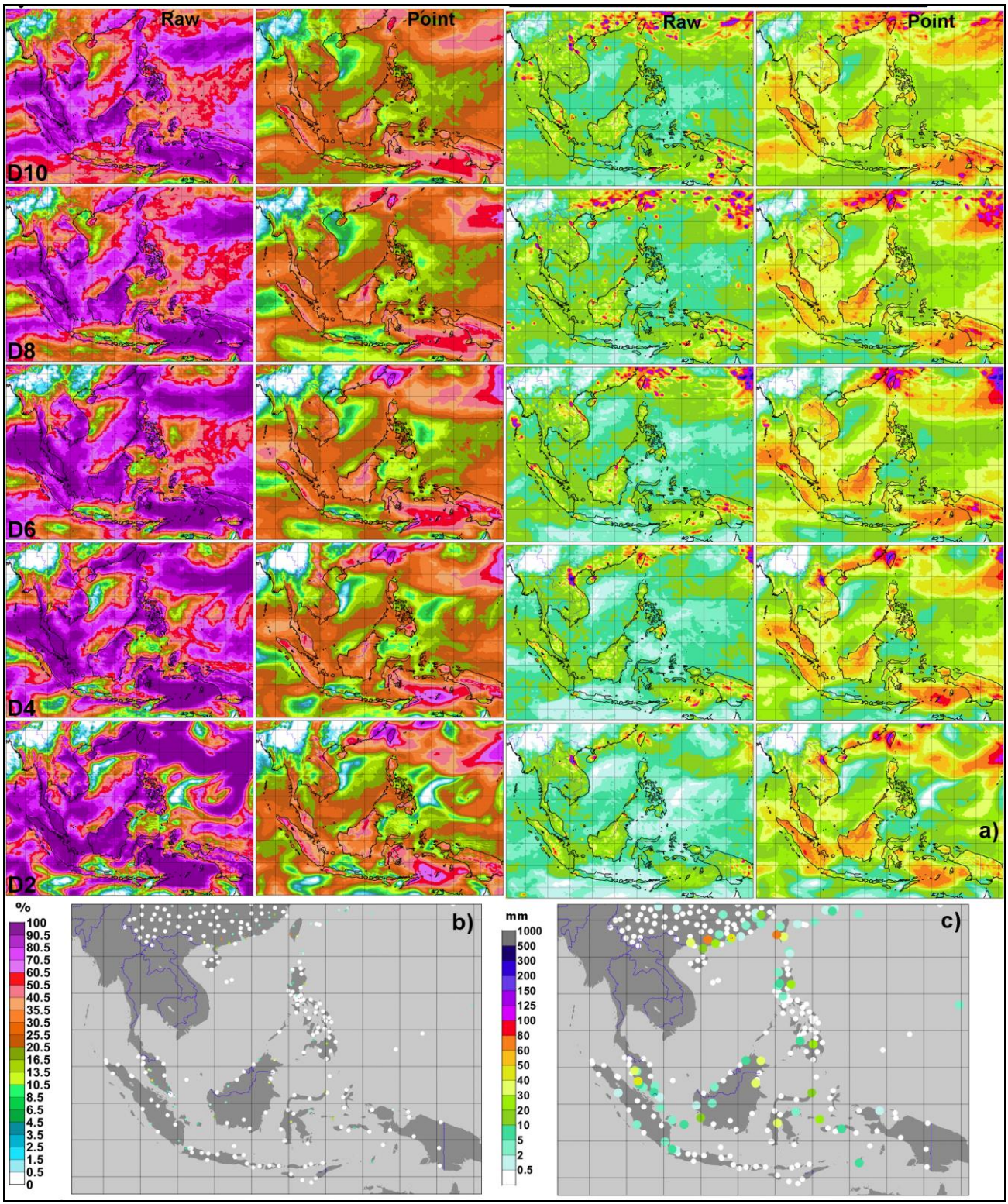

**Extended Data Fig. 3**: **Temporal progression of rainfall forecasts for one 12h period, for a tropical region, with verification**. As Extended Data Fig. 2 but for a valid period of 06 to 18UTC on 20th May 2019, and for forecast start times of 00UTC on 11th, 13th, 15th, 17th, 19th May (top to bottom rows on (a)).

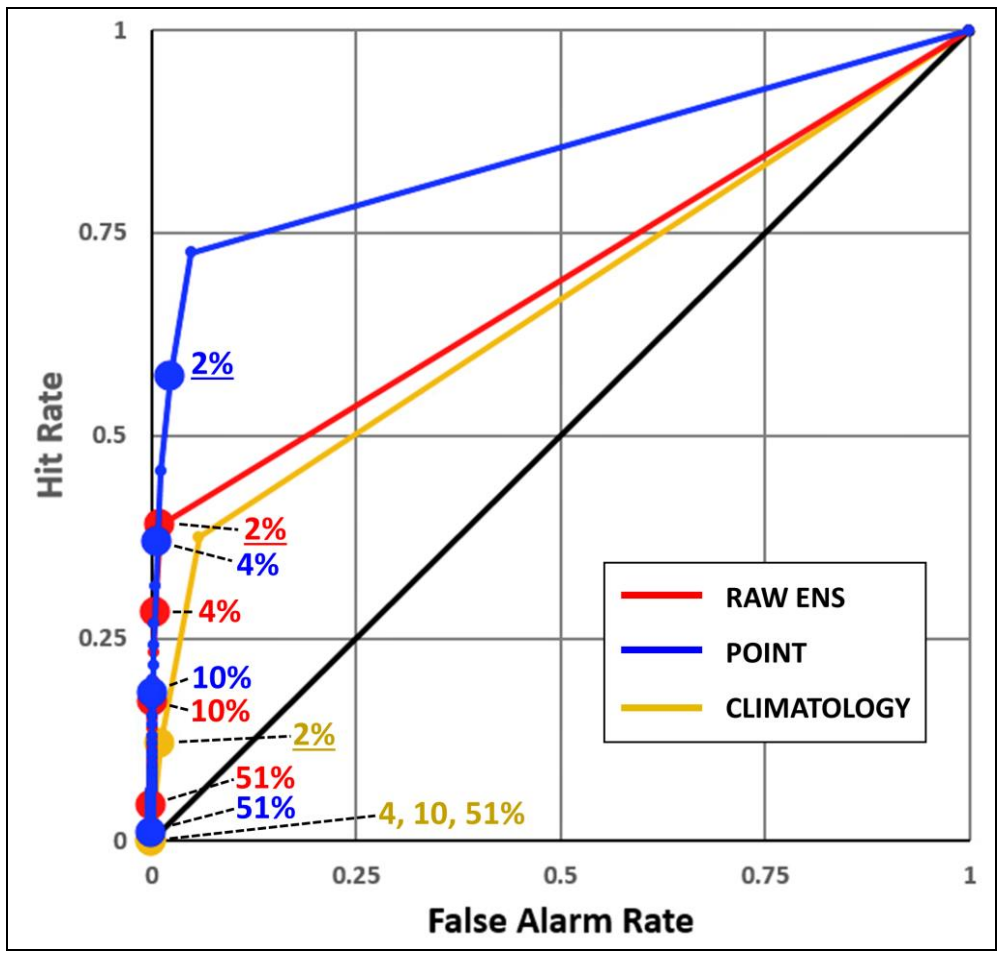

**Extended Data Fig. 4**: ROC curves from verification of one year of global raingauge observations. For forecasts of 12h rainfall, for 12-24h lead time, by raw ensemble/point rainfall/climatology (Red/blue/ yellow respectively). Spots are shown at integer percentiles; 1,2,..99 for point rainfall and climatology and wherever defined for the raw ensembles (51 members). Large spots signify probabilities of 2, 4, 10 and 51% as labelled. For climatology: the highest probability of >50mm/12h in our global data is 4%, so for >4% there are no cases to consider.

## Acknowledgements
D. Richardson provided helpful feedback on earlier drafts of this manuscript, particularly regarding verification aspects. A. Bonet assisted with the operationalization of ecPoint. M.A.O. Køltzow provided verifying data for the case in Fig. 5.

## Contributions
T.H. invented ecPoint, investigated and selected case studies, calculated climatological forecast skill and prepared the manuscript. F.P. investigated governing variable utility, coded up the PP system, performed ecPoint calibration and verification, and contributed to drafting of the manuscript.

## Competing interests
The authors declare no competing interests.